\documentclass[aps,prl,twocolumn,showpacs,superscriptaddress,showkeys,linenumbers]{revtex4-1}
\usepackage[T1]{fontenc}        
\usepackage{palatino}           
\usepackage{amsmath}            
\usepackage{amssymb}            
\usepackage{wasysym }           
\usepackage{graphicx}           
\usepackage{rotating}
\usepackage{booktabs}
\usepackage{pslatex}
\usepackage{epsfig}
\usepackage{epsf}
\usepackage[usenames]{xcolor}
\usepackage{mathrsfs}           
\newcommand{\sfrac}[2]{\frac{#1}{#2}}
\usepackage{caption}
\newcommand{\V}[1]{\mathbf{#1}}

\newcommand{\rem}[1]{}          
\renewcommand{\deg}{$^\circ$}

\begin{document}

\title{Magnon breakdown in a two dimensional triangular lattice Heisenberg antiferromagnet of multiferroic LuMnO$_3$}
\author{Joosung Oh}
\author{Manh Duc Le}
\author{Jaehong Jeong}
\affiliation{Center for Correlated Electron Systems, Institute for Basic Science (IBS), Seoul National University, Seoul 151-747, Korea}
\affiliation{FPRD, Department of Physics \& Astronomy, Seoul National University, Seoul 151-747, Korea}
\author{Jung-hyun Lee}
\affiliation{Center for Strongly Correlated Materials Research, Seoul National University, Seoul 151-747, Korea}
\affiliation{Department of Physics, Sungkyunkwan University, Suwon 440-746, Korea}
\author{Hyungje Woo}
\affiliation{ISIS Facility, STFC Rutherford Appleton Laboratory, Oxfordshire OX11 0QX, United Kingdom}
\affiliation{Department of Physics, Brookhaven National Laboratory, Upton, N.Y. 11973, USA}
\author{Wan-Young Song}
\affiliation{Center for Strongly Correlated Materials Research, Seoul National University, Seoul 151-747, Korea}
\affiliation{Department of Physics, Sungkyunkwan University, Suwon 440-746, Korea}
\author{T. G. Perring}
\affiliation{ISIS Facility, STFC Rutherford Appleton Laboratory, Oxfordshire OX11 0QX, United Kingdom}
\author{W. J. L. Buyers}
\affiliation{Chalk River Laboratories, National Research Council, Chalk River, Ontario K0J 1J0, Canada}
\author{S-W. Cheong}
\affiliation{Rutgers Center for Emergent Materials and Department of Physics and Astronomy, Rutgers University, Piscataway New Jersey 08854, USA}
\author{Je-Geun Park}
\email{jgpark10@snu.ac.kr}
\affiliation{Center for Correlated Electron Systems, Institute for Basic Science (IBS), Seoul National University, Seoul 151-747, Korea}
\affiliation{FPRD, Department of Physics \& Astronomy, Seoul National University, Seoul 151-747, Korea}
\affiliation{Center for Strongly Correlated Materials Research, Seoul National University, Seoul 151-747, Korea}

\date{\today}


\begin{abstract}

The breakdown of magnons, the quasiparticles of magnetic systems, has rarely been seen. By using an inelastic neutron scattering technique we report the observation of spontaneous magnon decay in multiferroic LuMnO$_3$, a simple two-dimensional Heisenberg triangular lattice antiferromagnet, with large spin, $S=2$. The origin of this rare phenomenon lies in the non-vanishing cubic interaction between magnons in the spin Hamiltonian arising from the noncollinear 120\deg~spin structure. We observed all three key features of the nonlinear effects as theoretically predicted: a roton-like minimum, a flat mode, and a linewidth broadening, in our inelastic neutron scattering measurements of single crystal LuMnO$_3$. Our results show that quasiparticles in a system hitherto thought of as ``classical'' can indeed break down.

\end{abstract}


\maketitle


The notion of a renormalized and stable quasiparticle, introduced by Landau for the Fermi liquid\cite{n1}, where the behavior of strongly interacting real particles is replaced by weakly interacting collective excitations or quasiparticles, is fundamental to modern theories of condensed matter physics. For example, an understanding of the electron quasiparticle dispersion is central to research in high temperature superconductors\cite{n2,n3,n4}. Despite the success of the theories based on stable quasiparticles, their breakdown has been predicted and indeed observed in some rare cases. The prime example is the breakup of electrons into spinons and holons in 1D quantum spin systems\cite{n5,n6}.

The magnon is the quasiparticle of magnetic systems with long-range order. Arguably the most detailed information on such systems, particularly on the interactions between magnetic moments, can be obtained by measuring the properties of magnons, such as their dispersion curve, for which inelastic neutron scattering is especially suited\cite{n7,n8}. However, just like the breakdown of electron quasiparticles in a 1D chain, magnons can break down under certain unique conditions, which has been observed in cases with $S=\sfrac{1}{2}$\cite{n10}.

Recently, spontaneous magnon decay has been proposed to occur even in more classical-like large spin systems\cite{n11,n12,n14,n15}. The essence of this theory is that in the 2D triangular lattice Heisenberg antiferromagnet (THAF) with a noncollinear ground state, the cubic terms in the expansion of the Holstein-Primakoff expression for the spin operators are not prohibited by symmetry, unlike for collinear magnetic order. The noncollinear order permits coupling between $S^z$ spin components along the moment direction on one sublattice with $S^{x,y}$ transverse components on other sublattices. The transverse (longitudinal) fluctuations include one- (two-) magnon terms so mixing these terms allows the decay of single magnons into two when kinematic constraints are met\cite{n14}. This coupling is also responsible for a $q$-dependent renormalization of the single magnon energies which results in a roton-like minimum in the dispersion and flattening of the top of the spectrum\cite{n12,n14}.

\begin{figure}
  \begin{center}
    \includegraphics[width=\columnwidth,viewport=21 458 588 824]{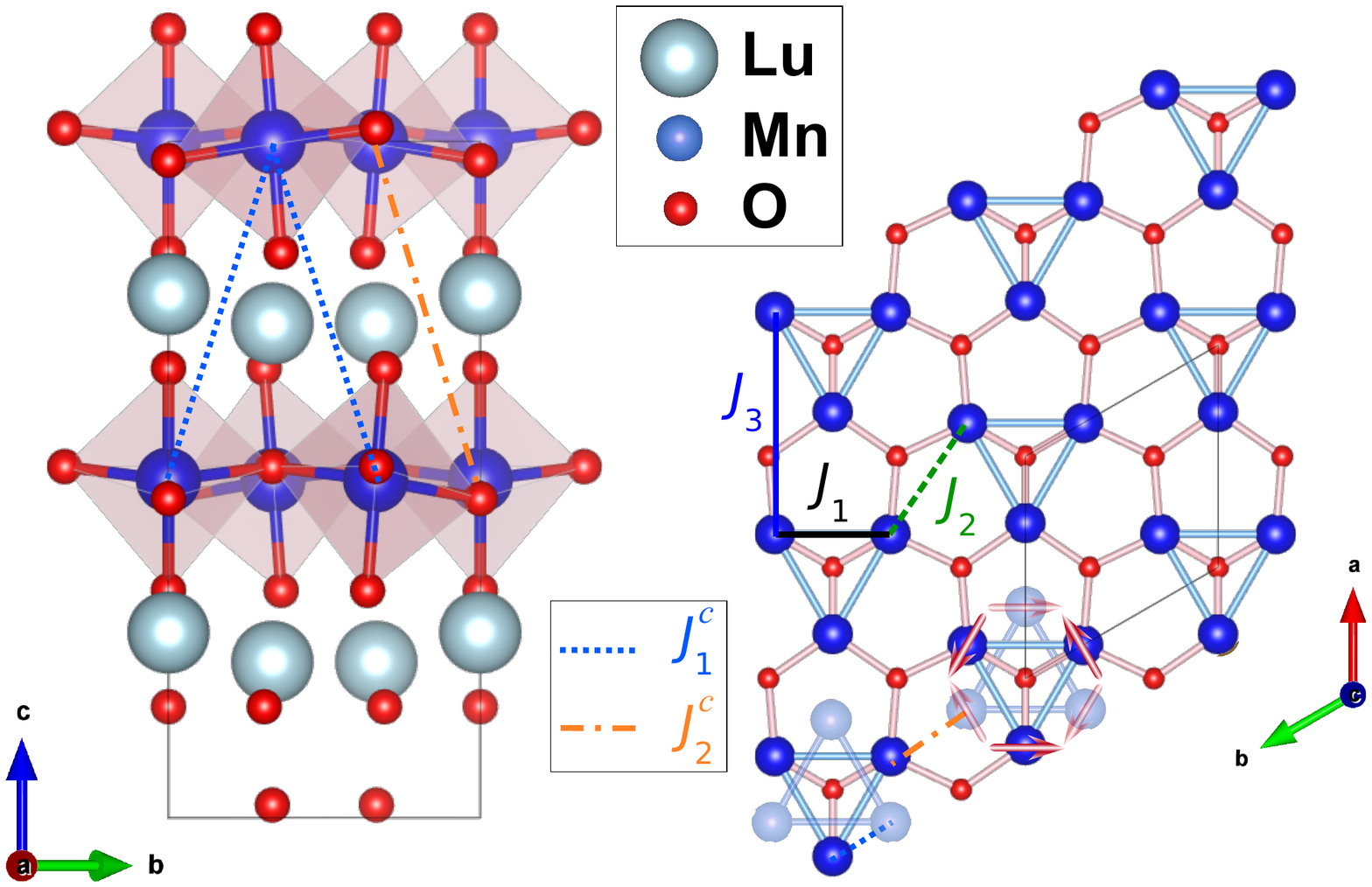}
    \caption{(Color online) The structure of LuMnO$_3$ showing the Mn-O plane (left), triangular lattice (right), the trimer units (light triangles), and exchange interactions considered in the spin Hamiltonian (thick lines).}
    \label{fig1} \end{center}
\end{figure}

\begin{figure*}
  \begin{center}
    \includegraphics[width=0.7\textwidth,viewport=19 225 594 586]{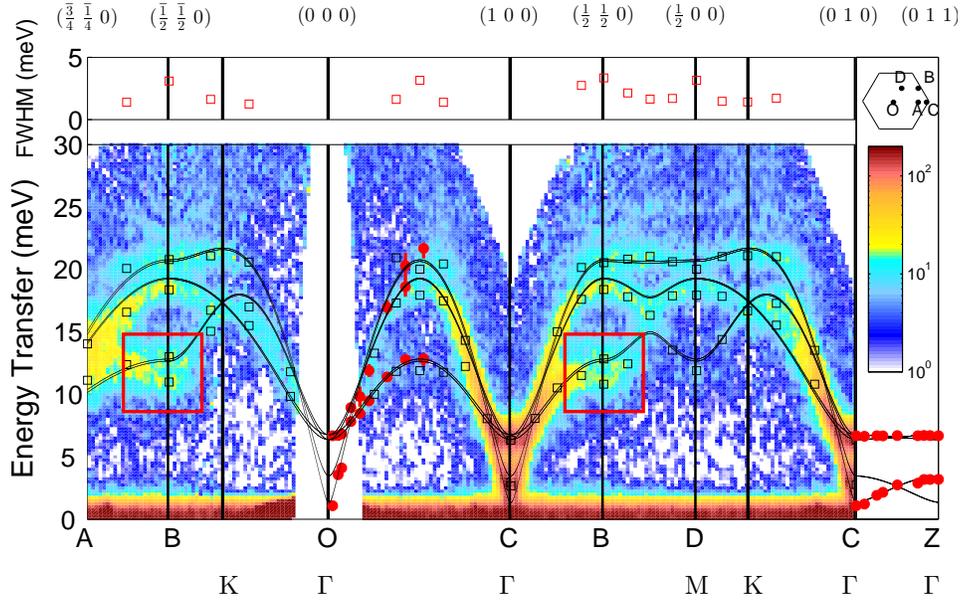}
    \caption{(Color online) Inelastic neutron scattering data along high symmetric directions: fitted peak positions from TAS data (\textcolor{red}{$\bullet$}), ToF data ($\square$ and the contour map), and the fitted dispersion (solid curves) calculated by linear spin wave theory. First Brillouin zone labels for the hexagonal unit cell (bottom text line) and triangular unit cell (line above) are also shown, together with a sketch of the triangular Brillouin zone (top right corner). The top panel shows the fitted FWHM of the 20~meV peaks from the ToF data, indicating broad peaks, possibly due to magnon decay, only near $(\frac{1}{2}\frac{1}{2}0)$ and $(\frac{1}{2}00)$.}
    \label{fig2} \end{center}
\end{figure*}

In this Letter we report direct experimental evidence of magnon breakdown in  LuMnO$_3$, which is a 2D THAF with a noncollinear 120\deg~structure and $S=2$. Our results demonstrate that although the overall features of the measured dispersion curves are consistent with linear spin wave theory, there are unmistakable signs of magnon breakdown exactly where the theory\cite{n18} predicts such highly unusual behavior to occur. 

LuMnO$_3$ forms in a layered structure with the $P6_3cm$ space group and belongs to the famous multiferroic hexagonal manganites\cite{n19,n20}. As an improper ferroelectric it undergoes a ferroelectric transition at 1050 K from centrosymmetric $P6_3mmc$ to the noncentrosymmetric $P6_3cm$. The origin of this ferroelectric transition was shown to be due to the buckling of the MnO$_5$ bipyramid and $pd$ hybridization\cite{n30,n31}, which also results in a trimerization of the 2D Mn triangular lattice\cite{n32}. Upon further cooling, the 2D Mn network undergoes an antiferromagnetic transition to the so-called 120\deg~structure\cite{n21,n40}. Below this transition, the Mn moments become involved in a very unusual spin-lattice coupling leading to a giant off-centering of the Mn position\cite{n22,n23}. At the same time, this off-centering gives rise to a very large volume reduction below T$_N$=90~K, where the anharmonic phonons are frozen so no thermal expansion is expected\cite{n24}.

In the antiferromagnetic phase, Mn moments form a distorted triangular lattice as shown in Fig.~\ref{fig1}. The spin dynamics of the Mn moments can be described by the following spin Hamiltonian:

{\small
\begin{gather} \nonumber
\mathcal{H} =
   -\mathcal{J}_1   \sum_{\mathrm{intra}} \V{S}_i\cdot\V{S}_j
   -\mathcal{J}_2   \sum_{\mathrm{inter}} \V{S}_i\cdot\V{S}_j
   -\mathcal{J}_3   \sum_{\mathrm{next\ nn}} \V{S}_i\cdot\V{S}_j
   -\mathcal{J}_1^c \sum_{\mathrm{out\ intra}} \V{S}_i\cdot\V{S}_j \\
   -\mathcal{J}_2^c \sum_{\mathrm{out\ inter}} \V{S}_i\cdot\V{S}_j 
   -D_1 \sum_i \left( \V{S}_i^z \right)^2
   -D_2 \sum_i \left( \V{n}\cdot\V{S}_i^z \right)^2
\end{gather}
}

\noindent where $\mathcal{J}_1$ ($\mathcal{J}_1^c$) and $\mathcal{J}_2$ ($\mathcal{J}_2^c$) are the intra- and inter-trimer in-plane (out-of-plane) exchange coupling respectively, $\mathcal{J}_3$ is the in-plane next nearest coupling, while $D_1$ and $D_2$ are magnetic anisotropy constants. The distinction between $\mathcal{J}_1$ and $\mathcal{J}_2$ arise from the off-centering of Mn $a$-axis displacement, $x$, below $T_N$\cite{n22}. This further doubles the number of allowed spin wave modes to six although they are nearly degenerate except near the $\Gamma$ point\cite{n50}.

\begin{figure}
  \begin{center}
    \includegraphics[width=\columnwidth,viewport=42 200 569 598]{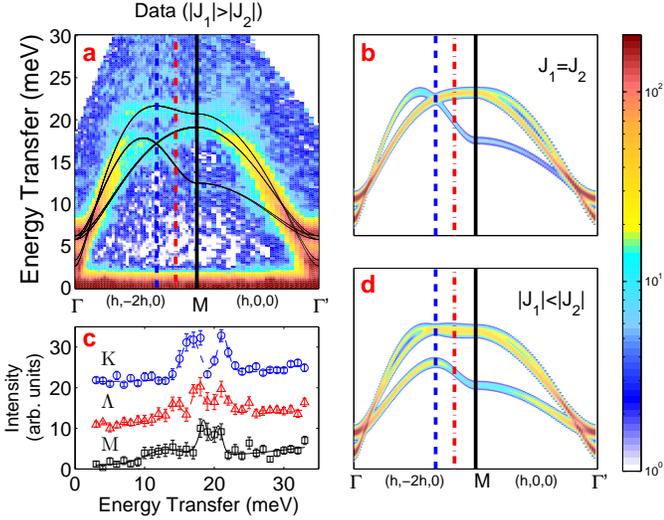}
    \caption{(Color online) Data ({\bf a}) and linear spin wave theory calculated neutron structure factors convoluted with an 0.8 meV Gaussian ({\bf b} and {\bf d}). {\bf c:} cuts along the vertical lines in the dispersion curves at the $M$, $(\sfrac{1}{2}00)$, (\textcolor{blue}{\rlap{-- --}~$\bigcirc$}~) and $K$, $(\sfrac{1}{2}\bar{\sfrac{1}{2}}0)$, (\rlap{-----}~$\square$~) wavevectors and in between, at $Q=(\sfrac{5}{12}\bar{\sfrac{1}{2}}0)$, (\textcolor{red}{\rlap{-$\cdot$-$\cdot$-}~$\triangle$}~).}
   \label{fig5} \end{center}
\end{figure}

The full dispersion curves of the spin waves of LuMnO$_3$, shown in Fig.~\ref{fig2}, were measured by inelastic neutron scattering on a single-crystal with total mass $\approx$3~g grown by using a commercial infrared mirror furnace (Crystal Systems, Japan). Measurements were carried out using the MAPS time-of-flight (ToF) spectrometer at the ISIS facility, UK, and the C5 triple axis spectrometer (TAS) at the Canadian Neutron Beam Center, Chalk River, Ontario. The incident energy was 40~meV for the ToF measurement, with the chopper speed set at 250~Hz in order to optimize the resolution, and the sample was mounted with the $(HHL)$ scattering plane horizontal and $k_i$ along $(001)$, such that the $(HK0)$ plane is imaged on the (vertical) detectors. A different horizontal scattering plane, $(H0L)$, was used for the TAS measurement, with the spectrometer configuration: 0.55\deg-PG(002)-0.48\deg-sample-0.55\deg-PG(002)-1.2\deg-detector, where the angles denote horizontal collimation and PG(002) is the Bragg reflection used for the monochromator and analyzer. In both cases the data was taken at 13~K, well below $T_N$.

The dispersion was calculated using standard methods with the best fit to the measured inelastic neutron spectra obtained by a minimal set of parameters: $\mathcal{J}_1$=-9 meV, $\mathcal{J}_2$=-1.4 meV, $\mathcal{J}_1^c$=-0.018 meV, $\mathcal{J}_2^c$=$\mathcal{J}_3$=0 meV, $D_1$=-0.28 meV, and $D_2$=0.006 meV. Except for some discrepancies related to the magnon decay discussed later, the key features of the measured spin waves are well captured by this model. 

\begin{figure*}
  \begin{center}
   \begin{tabular}{@{\extracolsep{\fill}}cc}
		\includegraphics[width=\textwidth,viewport=25 175 594 489]{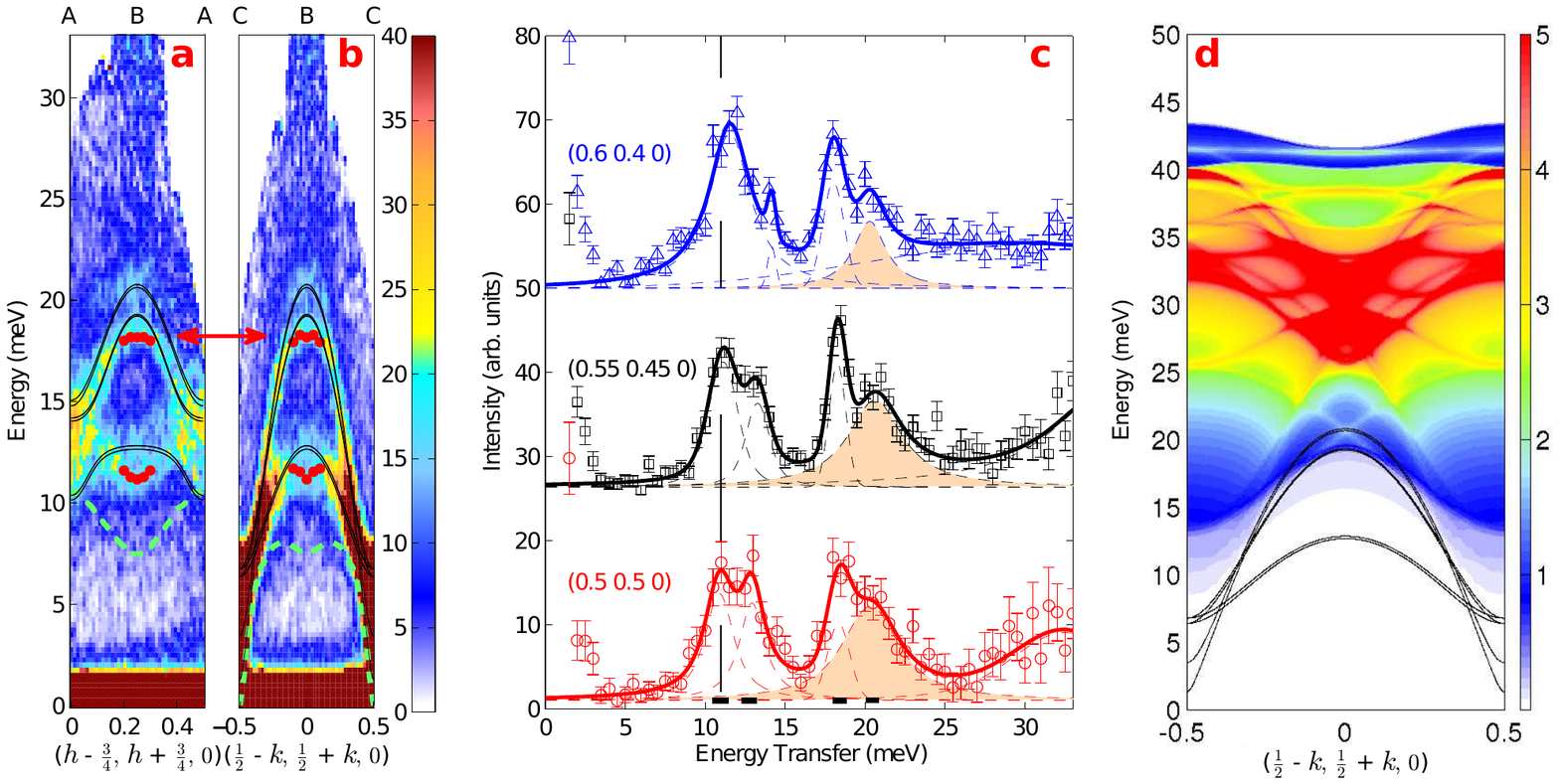}
   \end{tabular}
    \caption{(Color online) Cuts near the roton minimum showing the three signatures of magnon decay: (i) the minimum in the dispersion of the lowest energy mode at $(\sfrac{1}{2}\sfrac{1}{2}0)$ (panels {\bf a} and {\bf b}), the flat dispersion of the higher energy mode at the same point, indicated by the arrows in panels {\bf a} and {\bf b}; and the anomalously broad width of the $\approx$20~meV mode in the cuts in panel {\bf c}. In panels {\bf a} and {\bf b}, points (\textcolor{red}{$\bullet$}) indicate the fitted peak positions from energy cuts through the data. In panel {\bf c}, solid lines at the bottom directly below the peak centers indicate the instrumental resolution width. Thin dashed lines indicate individual fitted Voigt peaks, whilst the solid line is their sum, and points are measured data. The very broad peak at $\approx$32~meV in {\bf c} is attributed to 2-magnon scattering. Panel {\bf d} shows the two-magnon density of states calculated using linear spin wave theory from the single magnon dispersion (solid lines).}
    \label{fig3} \end{center}
\end{figure*}

The fits yield $|\mathcal{J}_1|>|\mathcal{J}_2|$, consistent with our high resolution neutron diffraction studies\cite{n22}, which found a large Mn off-centering distortion below $T_N$, resulting in two of the six nearest neighbor Mn-Mn bonds (corresponding to $\mathcal{J}_1$) becoming shorter than the others. This contrasts with previously reported TAS spin wave measurements\cite{n25} which suggested the opposite. In particular, the authors reported only two peaks at the $M$ point $(\sfrac{1}{2}00)$, which is only consistent with the case of $|\mathcal{J}_1| \leq |\mathcal{J}_2|$, as shown in Fig.~\ref{fig5}. Our data show three modes at $M$ and a mode crossing at $K$ which may only be explained by $|\mathcal{J}_1| > |\mathcal{J}_2|$.

The large ratio $\mathcal{J}_1/\mathcal{J}_2\approx$6.4, albeit within the stability limit of the long range 120\deg~structure, unlike in LiVO$_2$\cite{pen}, is unexpected. In terms of the spin wave dispersion, it is required by the large gap between two upper spin wave modes, which is degenerate when $\mathcal{J}_1=\mathcal{J}_2$. A ferromagnetic next nearest neighbor interaction has the same effect, permitting a lower $\mathcal{J}_1/\mathcal{J}_2$. However, this also decreases the energy of the spin waves at $\Gamma$, requiring a higher single ion anisotropy to compensate. The best fit in this case was with $\mathcal{J}_1$=-6.4 meV, $\mathcal{J}_2$=-1.3 meV, $\mathcal{J}_3$=0.15 meV, $\mathcal{J}_1^c$=0.009 meV, $\mathcal{J}_2^c$=-0.009, $D_1$=-0.5 meV and $D_2$=0.009 meV, yielding $\mathcal{J}_1/\mathcal{J}_2\approx$5. However, we found no improvement in fit quality by including the $\mathcal{J}_3$ term, with $\epsilon=\sfrac{1}{N}\sum_i|E_i^{\mathrm{meas}}-E_i^{\mathrm{calc}}|$=1.19 compared to $\epsilon$=1.17 for the case $\mathcal{J}_3=0$. Furthermore, such a large $\mathcal{J}_3$=0.15 meV may not be realistic.

Physically, we may relate the $\mathcal{J}_1/\mathcal{J}_2$ ratio to the frustration parameter $\theta_{\mathrm{CW}}/T_N$, since in a mean field model, $\theta_{\mathrm{CW}}\propto\sum \mathcal{J}_{ij}$ but $T_N$ is proportional to the average of the exchanges. As $\theta_{\mathrm{CW}}/T_N\approx$10 in LuMnO$_3$ we may expect $\mathcal{J}_1/\mathcal{J}_2$ to be large. Indeed using a Monte Carlo model\cite{kawamura} with our exchange parameters yields $T_N^{\mathrm{MC}}=0.31S^2\bar{\mathcal{J}}/k_B$=56~K. Together with the mean field result $\theta_{\mathrm{CW}}^{\mathrm{MF}}=\sfrac{1}{3k_B}S(S+1)\sum_{ij}\mathcal{J}_{ij}=$550~K, these estimates are not qualitatively dissimilar to the measured values, $T_N=$90~K and $\theta_{\mathrm{CW}}\approx$800~K. However, ab initio calculations~\cite{solovyev} found a much lower $\mathcal{J}_1/\mathcal{J}_2\approx$1.2. Furthermore, it is curious that a larger Mn displacement in YMnO$_3$\cite{n22} gives a smaller ratio $\approx$1.7\cite{n29} compared with LuMnO$_3$, which may be related to the nature of the Mn displacements: In LuMnO$_3$, the distortion creates trimers, whereas in YMnO$_3$ a connected Kagom\'e-like network is formed. Finally, another possibility is that the $\mathcal{J}$ values obtained from linear spin wave theory may be changed by taking into account terms for magnon decay.

A closer inspection of the experimental spin wave dispersion curve reveals further interesting discrepancies, which cannot be explained by the linear spin wave calculations. The most notable discrepancy is seen near $(\sfrac{1}{2}\sfrac{1}{2}0)$ (labeled B in the single sublattice triangular Brillouin zone), where the experimental dispersion curve not only deviates from the theoretical results but also shows a minimum (see the region marked by the box in Fig.~\ref{fig2}). Surprisingly, this minimum occurs exactly at the same point where nonlinear spin wave theory predicts a roton-like minimum\cite{n12,n14}. Interestingly enough, a similar roton-like minimum was observed in an $S=\sfrac{1}{2}$\cite{n26} quantum spin liquid. 

In order to demonstrate this connection with the theoretical predictions further, we have plotted an enlarged view of the spin waves near $(\sfrac{1}{2}\sfrac{1}{2}0)$ together with the linear spin wave theory calculations (solid lines) in Fig.~\ref{fig3}. The thick dashed lines in panels a and b are taken from a series expansion calculation of the nonlinear spin wave dispersion\cite{n12} for an ideal triangular lattice with $S=\sfrac{1}{2}$ after adjusting the overall $\mathcal{J}$ value to 13.2 meV in order to match the spin wave energies of LuMnO$_3$. We note that the $S=\sfrac{1}{2}$ theoretical calculations show a large quantum renormalization due to mode repulsion between the two-magnon continuum and the single-magnon dispersion, which is expected to be much weaker for the current $S=2$ case, and thus accounts for the apparent downward shift of the calculated curve compared to our measurements. Moreover, as indicated by the red arrows in our data, the experimental spin wave becomes considerably flattened around $(\sfrac{1}{2}\sfrac{1}{2}0)$ as predicted from the nonlinear spin wave theory\cite{n15}. The downward shift at this flat mode is about 5\% of the linear spin wave energy. Note that it has been predicted to be 8\% for $S=\sfrac{3}{2}$\cite{mourigal}.

In addition to the roton-like minimum and the flat mode, the decay of a single magnon into two magnons is also predicted by the nonlinear spin wave theory. In fact, our results show such line broadenings near $(\sfrac{1}{2}\sfrac{1}{2}0)$ and $(\sfrac{1}{2}00)$, as shown in the top panel of Fig.~\ref{fig2} by the larger full width at half maximum (FWHM) of the fitted peaks from energy cuts to the data. Fig.~\ref{fig3}c shows such cuts around $(\sfrac{1}{2}\sfrac{1}{2}0)$ where the highest energy mode is several times broader than the instrument resolution whilst the three other branches have FWHM similar to the instrument resolution. The signal at higher energy transfer is likely to be caused by two magnon scattering\cite{mourigal}. Similar scattering at high energies was also observed in earlier measurements on YMnO$_3$\cite{n28}.

Furthermore, this observation of magnon decay is consistent with the calculated two magnon density of states in Fig.~\ref{fig3}d\cite{twomagnon} which show that the top of the single magnon dispersion coincides with a line of strong two-magnon density of states permitting many decay channels. This may explain the large energy linewidth observed in Fig.~\ref{fig3}c. Together with the roton-like minimum and flat mode, this constitutes convincing experimental evidence that cubic and higher order terms in the bosonisation of the spin operators, neglected in linear spin wave theory, are important in LuMnO$_3$. We note that a roton-like minimum, but not the other two features, was reported previously in $\alpha$-CaCr$_2$O$_4$\cite{n27}. 
The presence of the cubic term in the spin Hamiltonian may also contribute to the observed reduction of the ordered moment ($\mu_{\mathrm{ord}}=3.3~\mu_B$/f.u.) compared with the ionic value of 4~$\mu_B$\cite{n14, n51}.

In conclusion, we have shown with LuMnO$_3$ that a 2D triangular lattice antiferromagnet with relatively large spin, $S=2$, exhibits all three key features of nonlinear quantum effects in its spin wave: a roton-like minimum, a flat dispersionless mode, and magnon decay. These nonlinear effects arise from the noncollinear spin structure, which in the case of LuMnO$_3$ is the 120\deg~structure, suggesting that nonlinear quantum effect may still be observed in systems closer to the classical limit. As there are many other triangular lattice antiferromagnets with a noncollinear ordered structure, we expect to see many more spin systems to exhibit such highly interesting effects.

\begin{acknowledgments}
We thank A. L. Chernyshev, M. E. Zhitomirsky, R. Coldea, T. J. Sato, Y. K. Bang, D. Khomskii, D. C. Peets, H. Jin and M. Mostovoy for helpful discussions. This work was supported by the Research Center Program of IBS (Institute for Basic Science) in Korea: Grant No. EM1203. Work at the CSCMR and SKKU was partly supported by the National Research Foundation of Korea (grant nos. KRF-2008-220-C00012 \& R17-2008-033-01000-0). The work at Rutgers University was supported by the DOE under Grant No. DE-FG02-07ER46382.
\end{acknowledgments}



\begin{thebibliography}{35}
\bibitem [{n50()}]{n50}%
  \BibitemOpen
  \href@noop {} {}\bibinfo {note} {Note that Miller indices and Brillouin zone
  labels in this work refer to the hexagonal unit cell, in contrast to the
  reduced triangular cell used in the theoretical works, which show only one of
  the three equivalent sublattices of the 120\deg~structure and is thus three
  times larger than the hexagonal cell. As a result, the theoretical
  calculations show only one of the three equivalent allowed spin wave modes of
  the triangular lattice. For comparison with the theory, Brilluoin zone labels
  for the reduced cell is also used.}\BibitemShut {Stop}%
\bibitem [{two()}]{twomagnon}%
  \BibitemOpen
  \href@noop {} {}\bibinfo {note} {The two magnon energies were determined from
  $E_2(\V{k})=\sum_{n,m}\left[E_{1,n}(\V{q})+E_{1,m}(\V{k}-\V{q})\right]$ where
  $E_{1,n}(\V{q})$ is the single magnon energy of the $n^{\mathrm{th}}$ branch
  at momentum transfer $\V{q}$ sampled over the full Brillouin zone. Fine step
  sizes: 0.00025 r.l.u. for the $(hk0)$ plane with $0<h<1$ \& $0<k<1$ and 0.001
  r.l.u. in the $(\sfrac{1}{2}-k,\sfrac{1}{2}+k,0)$ direction, was used to
  sample the 1st Brillouin zone and energy bins of 0.01 meV to calculate the
  density of states.}\BibitemShut {Stop}%
\end{thebibliography}

%


\end{document}